\documentclass[aps,prb,epsf,twocolumn,showpacs]{revtex4-2}

\usepackage{CJK,amsmath,amssymb,graphics,epsfig,epstopdf,color,verbatim,ulem,braket,tabularx,float}
\usepackage[colorlinks,linkcolor=blue,citecolor=blue,urlcolor=blue]{hyperref}
\usepackage{graphicx}
\usepackage{amsmath}

\begin{document}
	
	\title{Spiral Phase and Phase Diagram of the $S$=1/2 XXZ Model on the Shastry-Sutherland Lattice}

	\author{Zhengpeng Yuan}
	\author{Muwei Wu}
	\author{Dao-Xin Yao}
	\email{yaodaox@mail.sysu.edu.cn}
	\author{Han-Qing Wu}
	\email{wuhanq3@mail.sysu.edu.cn}
	\affiliation{Institute of Neutron Science and Technology, \\Guangdong Provincial Key Laboratory of Magnetoelectric Physics and Devices, \\ 
    State Key Laboratory of Optoelectronic Materials and Technologies,\\
    \mbox{School of Physics, Sun Yat-sen University, Guangzhou, 510275, China.}}
	
	\date{\today}
	
	\begin{abstract}
        We investigate the ground-state phase diagram of the $S$=1/2 XXZ model on the two-dimensional Shastry-Sutherland lattice using exact diagonalization (ED), density-matrix renormalization group (DMRG), and cluster mean-field theory (CMFT) with DMRG as a solver. In the isotropic case ($\Delta=1$), CMFT results reveal an intermediate empty plaquette (EP) phase that has a lower energy than the full plaquette (FP) phase. However, due to mean-field artifacts, CMFT alone is not suitable for accurately determining phase boundaries. Therefore, we combined three methods to map out the reliable phase diagram. Our calculations show that the EP phase narrows as $\Delta$ deviates from unity and eventually vanishes. More importantly, we identify a spiral phase at small $\Delta$, which has not been reported in previous studies. This phase is clearly captured by DMRG simulations on long cylindrical geometries. The competition between the EP, spiral, and $xy$-AFM phases near their boundaries provides a plausible explanation for the emergent spin-liquid-like behavior in RE$_2$Be$_2$GeO$_2$, while shedding new light on the role of XXZ anisotropy in the Shastry-Sutherland XXZ model.
	\end{abstract}
	
	
	\maketitle
	
\section{Introduction}

Initially proposed as a theoretical model, the Shastry-Sutherland-lattice (SSL) Heisenberg model has emerged as an important frustrated quantum spin system since its experimental realization in the compound SrCu$_2$(BO$_3$)$_2$ \cite{SRIRAMSHASTRY19811069,PhysRevLett.82.3168,PhysRevLett.82.3701,Shin_Miyahara_2003,doi:10.1143/JPSJ.76.073710,doi:10.1073/pnas.1114464109,PhysRevLett.110.067210,PhysRevLett.113.067201,Haravifard2016,zayed2017,PhysRevLett.124.206602,Jimenez2021,Shi2022,Cui:2023jzi,Guo2025}. This model and its related material systems have spurred decades of combined theoretical, numerical, and experimental research. A concerted effort has been aimed at mapping its comprehensive phase diagram and understanding the nature of its ground state and excitations. The Heisenberg model on this lattice exhibits a quantum phase transition tuned by the ratio of inter-dimer to intra-dimer interaction, denoted as $g=J/J^\prime$. A small $g$ stabilizes the dimer phase, while a large $g$ leads to the antiferromagnetic (AFM) phase. At intermediate values of $g$, the prevailing consensus holds that the system hosts a plaquette phase\cite{PhysRevLett.84.4461,PhysRevB.64.134407,PhysRevB.66.014401,lou2012studyshastrysutherlandmodel,PhysRevB.87.115144,PhysRevB.92.094440,PhysRevX.9.041037,PhysRevB.100.140413,PhysRevB.105.L041115,PhysRevB.105.L060409,1655430696868-504655297,PhysRevB.107.L220408,PhysRevLett.131.116702,PhysRevLett.133.026502,PhysRevB.110.214410,chen2024spinexcitationsshastrysutherlandmodel,PhysRevB.111.134411,corboz2025quantumspinliquidphase}.

However, the nature of the phase transition between the plaquette phase and the AFM phase exhibits contradictory results when probed by different methods and research teams. Several numerical studies have revealed a continuous phase transition between the empty plaquette (EP) phase and AFM phase \cite{PhysRevB.66.014401,PhysRevX.9.041037,PhysRevLett.84.4461,PhysRevB.64.134407,PhysRevLett.133.026502}. Due to the fact that the AFM phase breaks SU(2) symmetry while the EP phase breaks some lattice symmetries, it is proposed that there is a deconfined quantum critical point (DQCP) between them \cite{PhysRevX.9.041037,doi:10.1126/science.1091806,PhysRevLett.133.026502, YCui2025}. However, other investigations have revealed discontinuities in the first derivative of the energy and hysteresis in the order parameter, indicating a weak first-order transition \cite{PhysRevB.87.115144,PhysRevB.107.L220408,chen2024spinexcitationsshastrysutherlandmodel, LChen2025}. In addition, some studies suggest that there may also be a Quantum Spin Liquid (QSL) phase between the plaquette and AFM phases \cite{PhysRevB.105.L041115,PhysRevB.105.L060409,1655430696868-504655297,PhysRevB.111.134411,corboz2025quantumspinliquidphase}.

\begin{figure}
  \centering
  \includegraphics[width=0.5\textwidth]{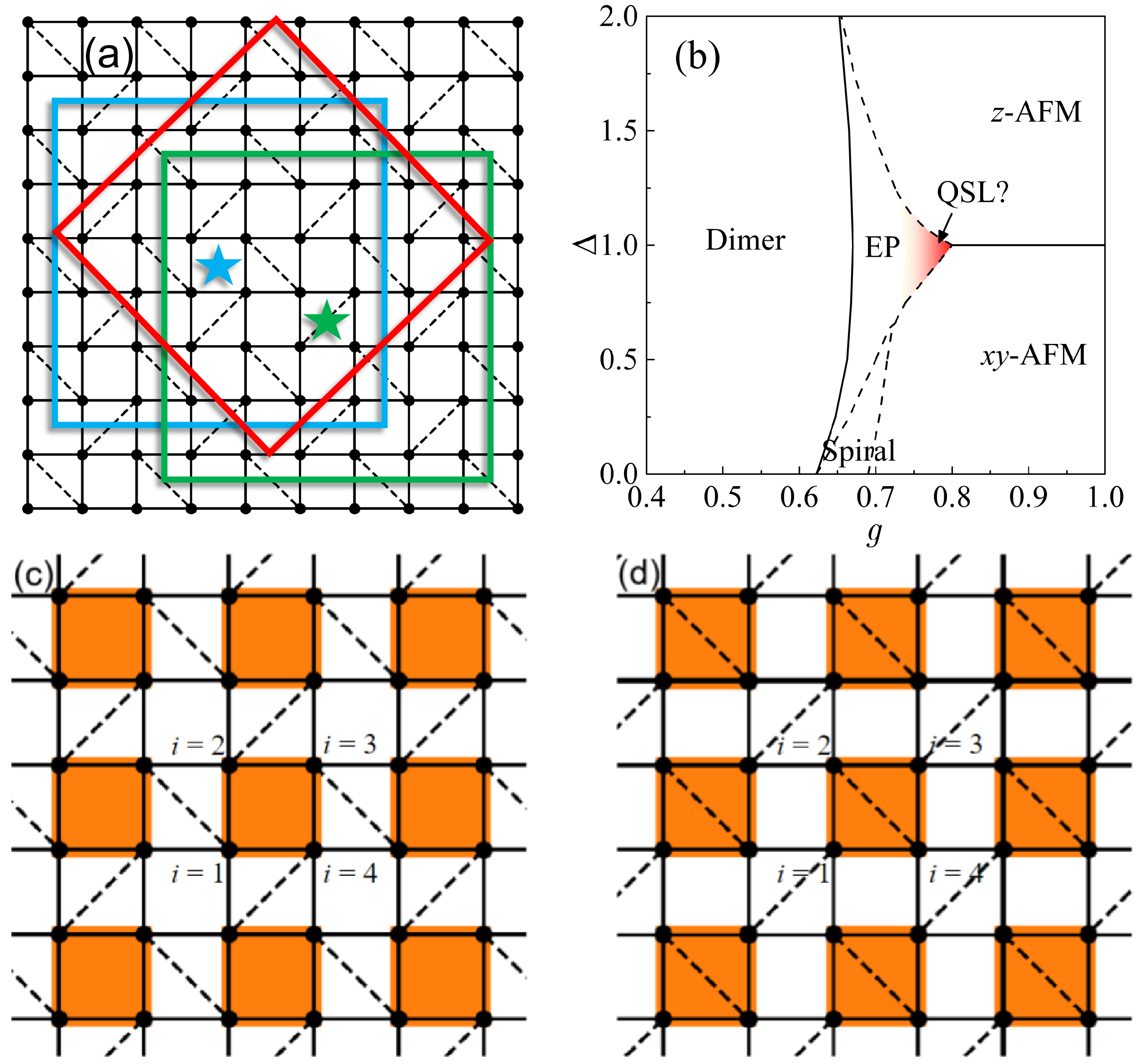}
  \caption{(a) Shastry-Sutherland lattice and three clusters used in the numerical calculations. Blue and green boxes represent $6\times6$ E- and F- type clusters, respectively, with stars of the same colors indicating their centers. The red box denotes the 32-site cluster used in ED calculations. (b) Phase diagram of the XXZ model on the Shastry-Sutherland lattice constructed from combined ED, CMFT, and DMRG results. Solid and dashed lines correspond to first-order and (possible) continuous phase transitions, respectively. The red gradient area in the figure indicates a possible spin-liquid region that is not excluded by our current results.
  (c,d) Schematic of empty plaquette (EP) and full plaquette (FP) states. Here, orange-shaded $2\times2$ plaquettes indicate plaquette-singlet patterns.}
  \label{fig:lattice}
\end{figure}
	
A recent experimental study on SrCu$_2$(BO$_3$)$_2$ has found that a first-order character of the plaquette-to-AFM phase transition weakens with increasing magnetic field, approaching the DQCP \cite{Cui:2023jzi}. Furthermore, most experimental results suggest that the intermediate phase is a full plaquette (FP) phase \cite{zayed2017,Jimenez2021,Cui:2023jzi}, unlike the theoretical EP phase. Interestingly, a more recent experiments have observed signals indicative of EP state under magnetic field and pressure\cite{cui2024plaquettesingletphasesshastrysutherlandcompound}. Some theoretical studies suggest that the full-plaquette state is a metastable state of the Heisenberg model on SSL and that the full-plaquette phase can be stabilized by modifying the model \cite{PhysRevB.83.140414,PhysRevB.100.140413,PhysRevB.107.L220408,LChen2025}. In addition to SrCu$_2$(BO$_3$)$_2$, certain rare-earth materials RE$_2$Be$_2$GeO$_7$ (RE = Pr, Nd, Gd-Yb), ErB$_4$ can also be described by magnetic SSL and are candidates for realizing some nonmagnetic phases like QSL \cite{doi:10.1021/acs.inorgchem.0c03131,PhysRevB.110.014412,PhysRevB.110.144445,li2024spinonsnewshastrysutherlandlattice,m3wx-4v6k}, spin-ice~\cite{ALiu2024}, “triplet” dimer phase~\cite{CLiu2025}. In these system, due to the strong spin-orbital effect, exchange anisotropy is key to understanding these materials \cite{PhysRevB.110.214410, CLiu2025}. While previous studies have outlined parts of the ground-state phase diagram for the XXZ model on the Shastry-Sutherland lattice\cite{PhysRevB.92.094440,PhysRevB.110.214410}, a comprehensive picture has remained elusive. By employing a combination of several numerical techniques, our work resolves the complete phase diagram and, most significantly, uncovers a new spiral phase near the regime of the XX model limit ($\Delta=0$). Our findings suggest that the interplay of the EP, spiral, and $xy$-AFM orders near their phase boundaries may account for part of the spin-liquid-like phenomenology in RE$_2$Be$_2$GeO$_2$ compounds.
	
In this paper, we investigate the $S$=1/2 Shastry-Sutherland lattice XXZ model using several numerical methods, including exact diagonalization (ED), cluster mean field theory (CMFT) \cite{DU2003387,PhysRevB.76.144420,PhysRevB.79.144427,PhysRevA.85.021601,PhysRevB.86.054516,PhysRevB.86.054516,PhysRevA.87.043619,PhysRevA.88.033624,Ren_2014,PhysRevLett.112.127203} and density matrix renormalization group (DMRG). In ED calculations, we use energy level spectroscopy technique to estimate the phase transition points. In CMFT, we directly measure order parameters to get the ground-state phase diagram, and illustrate the characteristics of the plaquette phase using bond energies of two types of cluster. The clusters used in the CMFT are up to 64-sites. In addition, we use DMRG calculations with cylinder geometries to obtain the phase diagram. Compared with ED and CMFT, we find a new spiral phase in the small $\Delta$ region. Our study made a key exploration into the ground-state phase diagram of XXZ model on Shastry-Sutherland lattice using several methods. Our finding is importance for understanding the Shastry-Sutherland lattice magnetic materials with XXZ anisotropy.
	
This paper is structured as follows. Section~\ref{sec:ModelandMethods} introduces the model and methods, presents different lattice geometries, and outlines the DMRG settings. Section~\ref{sec:Numericalresults} presents numerical simulation results for different methods, characterizing the ground-state phases and phase boundaries. And finally we give a summary in Section~\ref{sec:Summary}. 
	
\section{MODEL AND METHODS}
\label{sec:ModelandMethods}
    
In this paper, we employ a topologically equivalent lattice resembling a square lattice, analogous to the Shastry-Sutherland lattice [see Fig.~\ref{fig:lattice}(a)]. The Hamiltonian of the $S$=1/2 XXZ model on the Shastry-Sutherland lattice is given by Equation~\ref{eq:Hmlt}, where \(\Delta\) represents the XXZ anisotropy. $\Delta=1$ and $\Delta=0$ corresponding to the isotropic case and the planar XX model, respectively. Meanwhile, as \(\Delta\) approaches infinity, the system towards the Ising limit. 
	\begin{equation}
    \label{eq:Hmlt}
		\begin{split}
			\hat{H} &= J \sum_{\langle i,j \rangle} \left[ S_i^x S_j^x + S_i^y S_j^y + \Delta S_i^z S_j^z \right] \\
			&\quad + J' \sum_{\langle\langle i,j \rangle\rangle} \left[ S_i^x S_j^x + S_i^y S_j^y + \Delta S_i^z S_j^z \right]
		\end{split}
	\end{equation}
where $J$ denotes the nearest-neighbor exchange interaction or inter-dimer interaction depicted by solid lines in Fig.~\ref{fig:lattice}(a), and $J^\prime$ denotes some next-nearest-neighbor diagonal exchange interaction or intra-dimer interaction shown by dashed lines in the same figure, forming an orthogonal pattern. To study the ground-state properties of this frustrated model, we employ three numerical methods: exact diagonalization (ED) , density matrix renomalization group (DMRG) and cluster mean field theory (CMFT) with DMRG as a solver. 

The ED method is chosen to determine the phase boundaries using energy-level crossings in a 32-site cluster with periodic boundary conditions. This finite-size cluster is illustrated by the red box in Fig.~\ref{fig:lattice} (a). In the ED calculations, we used $U(1)$ symmetry with magnetization $M_z=0,\pm 1, \pm2, \cdots$, spin-flip symmetry with eigenvalue $Z=\pm 1$, and translation symmetry $\hat{T}^{\vec{r}} |\psi(\vec{k})\rangle=e^{-i\vec{k}\dot \vec{r}}|\psi(\vec{k})\rangle$ to perform the diagonalization of blocks. 

In the CMFT calculations, we set a self-consistent Weiss field on the boundary sites. All the correlations within the clusters are fully considered, while inter-cluster correlations are treated in a mean-field manner. Specifically, for the exchange interaction bonds between the clusters, the boundary interactions are replaced by mean-field terms, namely \(\langle S_i^\alpha \rangle S_j^\alpha \), as shown in Eq.~\ref{eq:Meanfield}. 
	\begin{equation}
    \label{eq:Meanfield}
		\begin{split}
			&S_i^x S_j^x + S_i^y S_j^y + \Delta S_i^z S_j^z \\
			\rightarrow&S_i^x \langle S_j^x \rangle + S_i^y \langle S_j^y \rangle + \Delta S_i^z \langle S_j^z \rangle\\
			+& \langle S_i^x \rangle S_j^x + \langle S_i^y \rangle S_j^y + \Delta \langle S_i^z \rangle S_j^z
		\end{split}
	\end{equation}
To retain the full space-group symmetry of the lattice, we use \(L \times L\) clusters (where \(L = 4, 6, 8\)) in the CMFT calculations. To further characterize the plaquette phase, we also employed two types of clusters corresponding to the E-type (centered on an empty plaquette) and the F-type (centered on a full plaquette). The illustrations of the E and F types are depicted by the blue and green boxes in Figure 1(a), where the pentagram of the corresponding color indicates the central position of the lattice type.

In the DMRG calculation, we employ the DMRG code from the ITensor library~\cite{ITensor, ITensor-r0.3}. For the DMRG calculations not serving as the solver of CMFT, we mainly use $L_y=6$ cylinders with $L_x=24$, keeping up to 6000 states and converging to a truncation error below $10^{-6}$. When DMRG is used as the solver within CMFT, we increase the maximum number of kept states to 7000, with a truncation error tolerance of \(1 \times 10^{-7}\) during sweeps and \(1 \times 10^{-6}\) for the energy convergence. Moreover, for cases with \(\Delta \geq 1\), $U(1)$ symmetry is imposed with an equal number of spin-up and spin-down states to accelerate the computations. To further improve convergence in the CMFT iterations, the successive over-relaxation (SOR) method is also applied.
	
To characterize the dimer, plaquette and N$\mathrm{\acute{e}}$el antiferromagnetic phases, we define the order parameters as follows:
\begin{equation}
	\begin{aligned}
		m_d^E =&\frac{1}{4}(\left| \braket{S_{\mathbf{r}_1}S_{\mathbf{r}_1-\mathbf{a}_1+\mathbf{b}_1}}-\braket{S_{\mathbf{r}_{2}}S_{\mathbf{r}_{2}-\mathbf{a}_1-\mathbf{b}_1}} \right| \\
        +&\left| \braket{S_{\mathbf{r}_2}S_{\mathbf{r}_2+\mathbf{a}_1+\mathbf{b}_1}}-\braket{S_{\mathbf{r}_{3}}S_{\mathbf{r}_{3}-\mathbf{a}_1+\mathbf{b}_1}} \right| \\
        +&\left| \braket{S_{\mathbf{r}_3}S_{\mathbf{r}_3+\mathbf{a}_1-\mathbf{b}_1}}-\braket{S_{\mathbf{r}_{4}}S_{\mathbf{r}_{4}+\mathbf{a}_1+\mathbf{b}_1}} \right| \\
        +&\left| \braket{S_{\mathbf{r}_4}S_{\mathbf{r}_4-\mathbf{a}_1-\mathbf{b}_1}}-\braket{S_{\mathbf{r}_{1}}S_{\mathbf{r}_{1}+\mathbf{a}_1-\mathbf{b}_1}} \right|) \\
        m_d^F =&(\left| \braket{S_{\mathbf{r}_2}S_{\mathbf{r}_4}>-<S_{\mathbf{r}_{1}}S_{\mathbf{r}_3}} \right|)\\
		m_p =& \frac{1}{4}\sum_{i=1}^4(\left| \braket{S_{\mathbf{r}_i}S_{\mathbf{r}_i+\mathbf{a}_1}}-\braket{S_{\mathbf{r}_i}S_{\mathbf{r}_i-\mathbf{a}_1}} \right|+\\
        &\left| \braket{S_{\mathbf{r}_i}S_{\mathbf{r}_i+\mathbf{b}_1}} - \braket{S_{\mathbf{r}_i}S_{\mathbf{r}_i-\mathbf{b}_1}} \right|) \\
		m_s =& \frac{1}{4} \sum_{i=1}^4 \sqrt{\langle S_i^z \rangle^2 + \langle S_i^x \rangle^2 + \langle S_i^y \rangle^2}
	\end{aligned}
\end{equation}
where $i$ represents the four lattice points at the center of the lattice (see Fig.~\ref{fig:lattice}(c) and ~\ref{fig:lattice}(d)). \(m_d^{E/F}\) denotes the dimer order parameter for E-type and F-type lattices, respectively. 

\section{Numerical results}
\label{sec:Numericalresults}

\subsection{32-site ED spectra}

\begin{figure}
  \centering
  \includegraphics[width=0.5\textwidth]{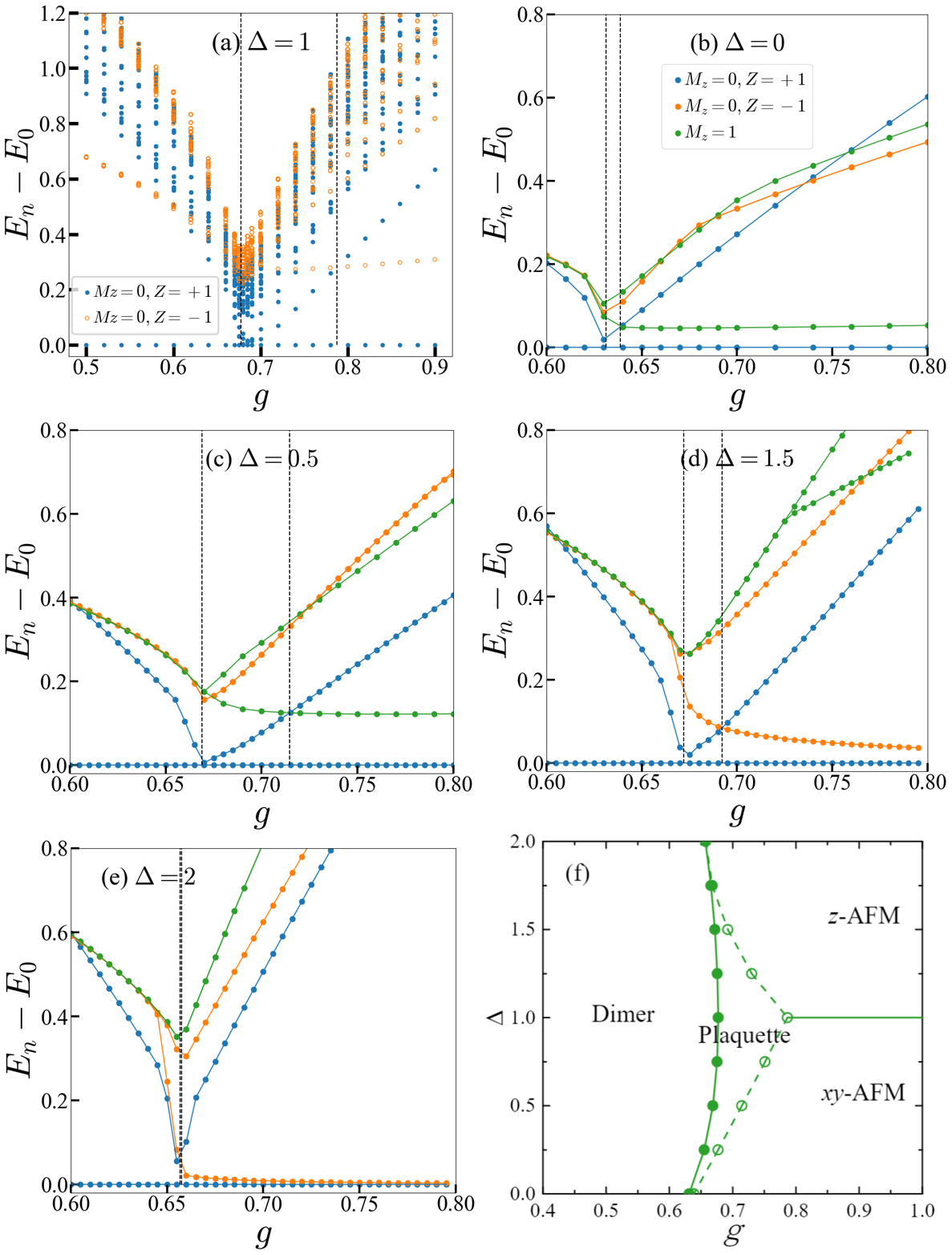}
  \caption{(a-e) Exact diagonalization (ED) energy spectra for a 32-site system at $\Delta=1, 0, 0.5, 1.5, 2$. The ED calculation uses symmetries including the total spin $M_z$ and spin-flip quantum number $Z$. Different colors represent the energy spectra from different symmetry sectors. Dashed vertical lines highlight the level crossing points indicating the quantum phase transitions. The phase diagram obtained by finite-size ED calculations is shown in (f).}
\label{fig:EDenergy}
\end{figure}

\begin{figure*}[htp]
  \centering
  \includegraphics[width=1.0\textwidth]{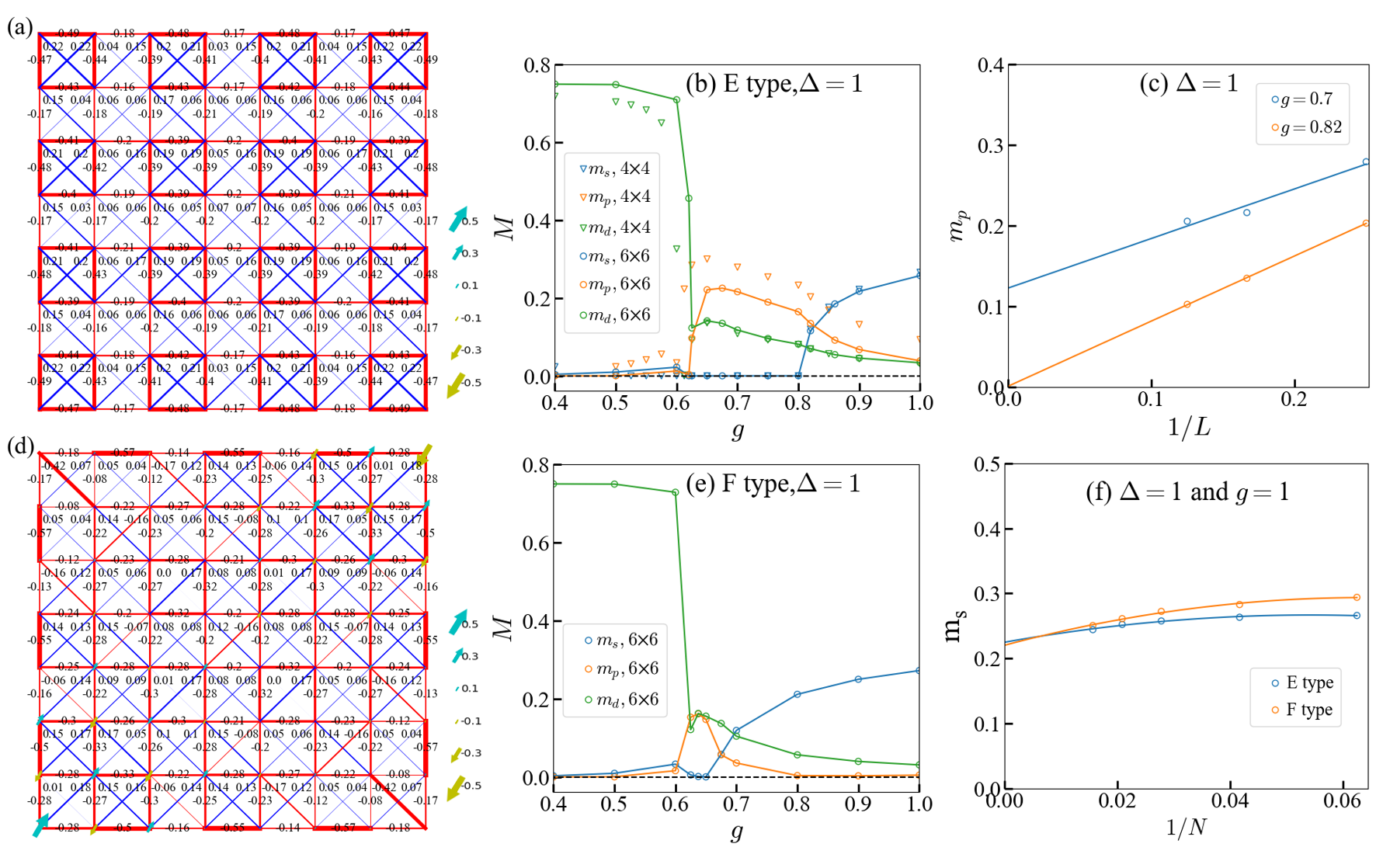}
  \caption{(a) and (d) illustrate the bond energy and magnetic order strength on each site for the E-type and F-type $8\times8$ clusters at $g=0.72$ and $g=0.66$, respectively. The evolution of dimer, plaquette, and AFM order parameters with $g$ is presented for the E-type cluster in (b) and for the F-type cluster in (e). Extrapolations of the order parameters are shown in (c) and (f): linear extrapolations of the plaquette order from the E-type cluster at $g=0.7$ and $0.82$ are displayed in (c), while second-order polynomial extrapolations of the AFM order obtained from both E-type and F-type clusters at $g=1$ are shown in (f).}
\label{fig:CMFTresults}
\end{figure*}

We first restudy the isotropic case with \(\Delta = 1\). Using exact diagonalization on a 32-site torus (periodic boundary condition), we tune $g$ and apply the level spectroscopy technique to identify quantum phase transitions~\cite{Suwa2016, LWang2018, Nomura2021, PhysRevB.105.L060409, 1655430696868-504655297, TTang2022, MWu2022, Wietek2024, YZheng2025}. Our results, illustrated in Fig.~\ref{fig:EDenergy}(a), reveal a ground state level crossing at \(g \approx 0.677\). This crossing is accompanied by a discontinuity in the first derivative of the energy with respect to \(g\), indicating a first-order phase transition. The Dimer phase and Plaquette VBS phase have different singlet pattern ground state. So it is natural to see the level crossing. This transition point estimated by 32-site ED aligns with previous literature reporting a phase transition point at \(g_{c1} \approx 0.675\) for dimer phase and plaquette phase \cite{PhysRevLett.84.4461,lou2012studyshastrysutherlandmodel,PhysRevB.87.115144,PhysRevX.9.041037,PhysRevB.107.L220408}.  
Furthermore, the plaquette phase exhibits a quasi-degenerate ground-state manifold of singlets in finite-size systems. In contrast, the antiferromagnetic phase at large $g$ breaks the $SU(2)$ symmetry, leading to an Anderson tower of states~\cite{Anderson1952, Claire2005, Wietek2017}: the ground state is a singlet, followed by a triplet as the first excited state. This difference suggests that a level crossing of the first excited state should occur between the two phases. Indeed, in the ED spectra, a crossing appears near $g_{c2}\approx 0.787$ between the first excited singlet ($S=0$) and the lowest triplet ($S=1$) states, which is consistent with previous studies using level spectroscopic technique~\cite{PhysRevB.105.L060409, 1655430696868-504655297}. The value is also in the range $0.77-0.828$ reported by employing various other methodologies \cite{PhysRevX.9.041037,PhysRevB.107.L220408,PhysRevLett.133.026502}, suggesting the potential existence of a continuous phase transition. Interestingly, near $g_{c1}$, four nearly degenerate singlet states emerge. This suggests a competition between the EP and FP states, with each phase contributing two quasi-degenerate states.

For the anisotropic XXZ Shastry–Sutherland model, we continue to employ energy level spectroscopy—as in the case of $\Delta=1$—to identify phase transitions, motivated by the distinct ground-state wavefunctions and low-energy excitation spectra exhibited by the dimer, EP, and antiferromagnetic (AFM) phases. Their transitions are marked by clear crossings in the low-lying energy spectrum, as shown in Fig.~\ref{fig:EDenergy}(b–e). Specifically, the transition point $g_{c1}$ from the dimer to the plaquette VBS phase is identified via ground-state level crossings within the same $M_z=0$ sector. Meanwhile, the transition from the plaquette to the AFM phase $g_{c2}$ is determined by the first-excited-state level crossing, using different $M_z$ sectors for $\Delta<1$ and spin-inversion parity or eigenvalues for $\Delta>1$. The resulting phase diagram from 32-site exact diagonalization is presented in Fig.~\ref{fig:EDenergy}(f). It seems that the plaquette VBS can extend to $\Delta=0$ case with a very narrow region. Combined with DMRG results, we will show that it is actually a finite-size effect.

\subsection{CMFT study on $L\times L$ clusters}

Next, we turn to the CMFT results. In the intermediate $g$ regime, for the E type cluster with an $8\times8$ size, the bond energy or nearest-neighbor spin correlation and the magnetic order $\langle \mathbf{S} \rangle$ at $g=0.76$ are depicted in Fig.~\ref{fig:CMFTresults}(a). A strong tetramer pattern centered on the empty plaquettes is evident, whereas the magnetic order is zero. For a direct comparison with the dimer and AFM order parameters, and to obtain the phase diagram, we plot the evolution of all order parameters obtained by $4\times 4$ and $6\times 6$ clusters as a function of $g$, which is shown in the Fig.~\ref{fig:CMFTresults}(b). As shown in the Fig.~\ref{fig:CMFTresults}(c) with $g=0.7$, EP order parameter $m_p$ extrapolates to a finite value with increasing system size, suggesting that the EP phase can exist stably in the thermodynamic limit. While at \(g = 0.82\), \(m_p\) extrapolates to zero , which means that plaquette order vanishes in the AFM phase. Starting from the dimer phase and increasing \(g\). A jump in the \(m_d\) order parameter indicates a first-order phase transition between the dimer and plaquette phases. The phase transition points are observed at \(g_{c1} \approx 0.602\) for 4×4 and \(g_{c1} \approx 0.6225\) for 6×6, which quite differ from ED and the \(g \approx 0.675\) reported in other literature. This discrepancy arises because improper mean-field treatment $\langle S_i \rangle S_j$ of strong dimer bond $\frac{1}{2}(\left|\uparrow\downarrow\right\rangle - \left|\downarrow\uparrow\right\rangle)$ used in the CMFT calculation, making the dimer state less competitive in energy and shifting the phase transition point to smaller $g$. The calculations for the current 4×4 and 6×6 clusters already show a trend of increasing phase transition positions. When the cluster size is sufficiently large, this transition point can be accurately determined. However, due to current computational constraints and convergence limitations, we are not yet able to accurately compute larger systems near this phase transition within the CMFT framework. 

Further increasing \(g\) leads the system continuously transition to the AFM phase. Importantly, the \(m_s\) order parameter shows little size dependence. For both 4×4 and 6×6 systems, the phase transition point is nearly fixed at \(g_{c2} \approx 0.8\), which is consistent with the range $[0.78, 0.826]$ found in other literature. The evidence therefore strongly supports the reliability of the phase transition point $g_{c2}$ identified by the CMFT method. Furthermore, the points where magnetic order vanishes, as reported in Refs.~\cite{Ren_2014} and \cite{PhysRevLett.112.127203} for other lattices, are also consistent with findings from other numerical techniques.
	
For the results obtained from $8\times8$ F-type cluster, at \(g = 0.66\), the bond energy and magnetic order $\langle \mathbf{S} \rangle$ of the ground state in the plaquette phase are shown in Fig.~\ref{fig:CMFTresults}(d). Compared to the E type of the same size, this configuration exhibits weak tetramer pattern at full plaquette centers with intradimer or diagonal interaction. In Fig.~\ref{fig:CMFTresults}(e), we present the order parameters for the F type cluster with $6\times6$ size (the full-plaquette state cannot be effectively detected in $4\times4$ cluster). In the order parameter diagram, the non-zero range of \(m_p\) is relatively narrow. This implies that the FP phase is only competitive against the dimer and AFM phases within a very narrow parameter range. A further comparison of the energies in their coexistence region obtained by CMFT reveals that the FP phase has a higher energy than the EP phase. Consequently, the EP phase is the only stable ground state in the thermodynamic limit. The intermediate phase is the EP phase. In contrast, the FP phase competes with the EP phase only in a narrow region near the dimer–EP phase transition. This is further supported by the four nearly degenerate states observed in the ED spectra in Fig.~\ref{fig:EDenergy}(a).

For the antiferromagnetic (AFM) phase region, we study the magnetic order at $g=1$, the extrapolation results of \(m_s\) for both E and F types are shown in Fig.~\ref{fig:CMFTresults}(f). The extrapolation values of \(m_s\) for the E and F types are approximately 0.224 and 0.220, respectively. These values are quite close as expected because the results of two type clusters should be equivalent in the thermodynamic limit. Compared to the magnetic order of the square lattice, the magnitude of the magnetic order is reduced due to frustration effects induced by intradimer or next-nearest-neighbor diagonal interaction.

\begin{figure}
  \centering
  \includegraphics[width=0.5\textwidth]{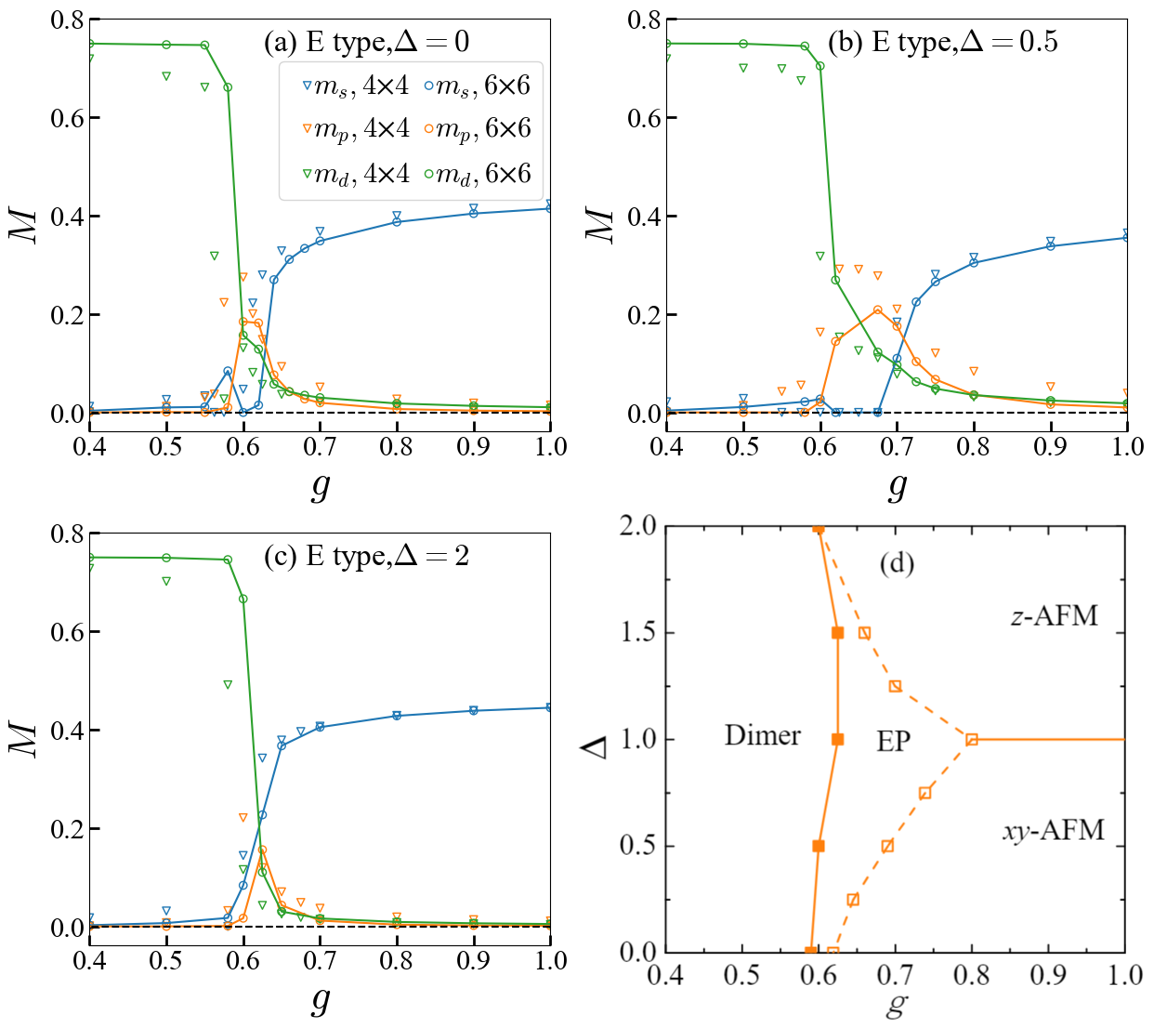}
  \caption{(a-c) The order parameter as a function of g for E-type structures when $\Delta = 0, 0.5, 2$ in sizes of 4$\times$4 and 6$\times$6. (d) Phase diagram obtained by collecting the phase boundaries calculated via CMFT for $\Delta = 0, 0.5, 1, 1.5, 2$ in size of 6$\times$6. Solid symbols and solid lines represent first-order phase transitions, while open symbols and dashed lines represent continuous phase transitions.}
\label{fig:CMFTOtherDelta}
\end{figure}
    
Similar to the isotropic case of $\Delta = 1$, we employ cluster mean-field theory (CMFT) with $4 \times 4$ and $6 \times 6$ E‑type clusters to obtain the phase diagram of the XXZ model. The dimer, plaquette, and AFM order parameters as functions of $g$ for different $\Delta$ are shown in Fig.~\ref{fig:CMFTOtherDelta}(a–c). Quantum phase transitions are identified by jumps in the dimer order parameter and the emergence of AFM order. The full $g$–$\Delta$ phase diagram, marked by yellow squares in Fig.~\ref{fig:CMFTOtherDelta}(d), reveals that the intermediate plaquette phase narrows with increasing anisotropy, consistent with the ED results. At $\Delta = 0$, CMFT based on the $6 \times 6$ cluster suggests the persistence of an intermediate plaquette phase; however, this is likely a finite‑cluster effect, which will be further examined using DMRG calculations later. For $\Delta > 1$, the intermediate plaquette phase vanishes between $\Delta = 1.5$ and $2$.

\begin{figure*}[htp]
  \centering
  \includegraphics[width=1.0\textwidth]{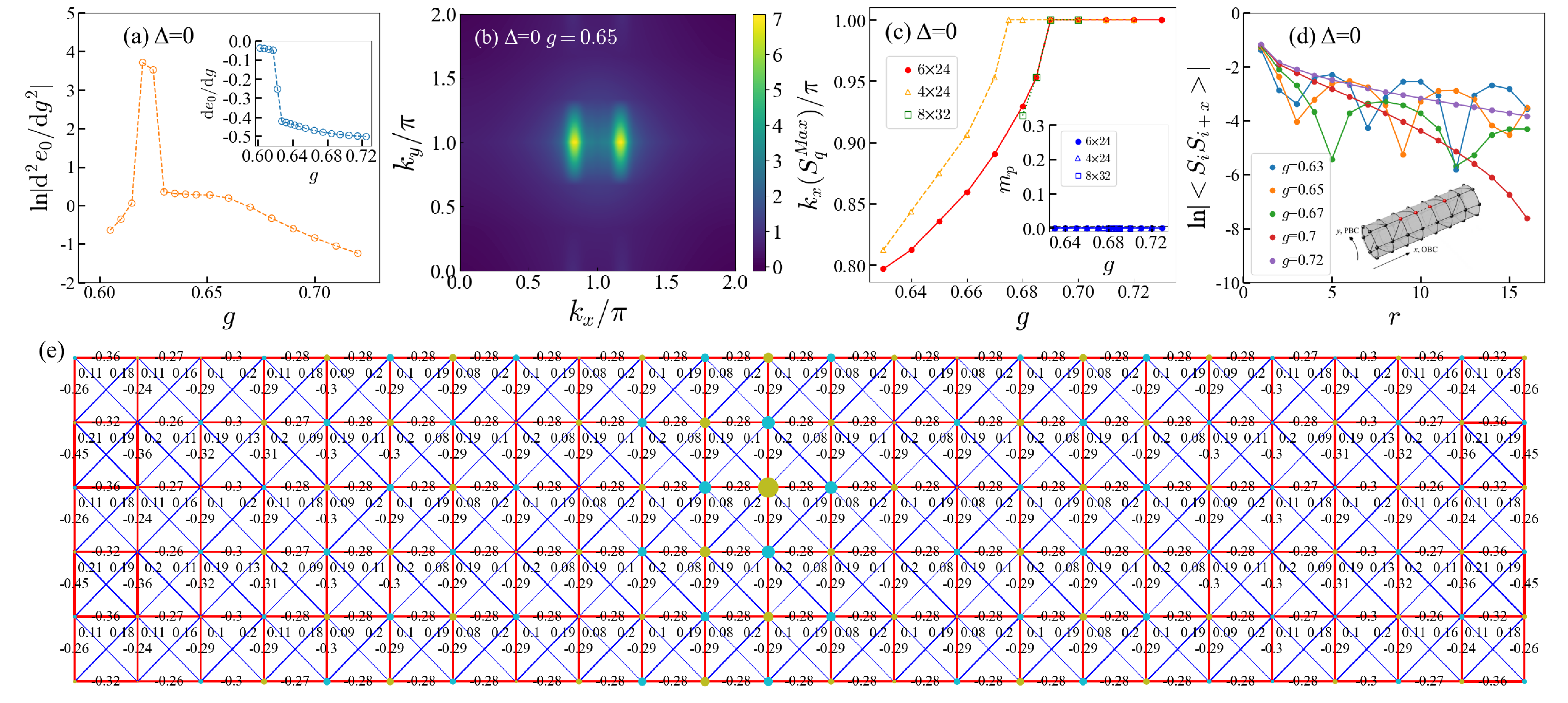}
  \caption{(a) Logarithm of the absolute second derivative of the ground-state energy as a function of $g$. The inset shows the corresponding first derivative. (b) Spin structure factor at $\Delta=0$ and $g=0.65$, exhibiting two magnetic Bragg peaks near $(\pi,\pi)$. (c) Evolution of the magnetic Bragg peak positions with $g$. While the $L_y=4$ cylinder exhibits finite-size effects, the $L_y=6$ and $L_y=8$ results at $g=0.68, 0.685, 0.69, 0.7$ (open squares) are very close, indicating weak finite-$L_y$ effects for $L_y=6$. The inset shows a negligible plaquette VBS order.
  (d) Real-space spin-spin correlation functions along the $x$ direction for various values of $g$. (e) Schematic representation of spin-spin correlations: between the reference site and all other sites, and the nearest-neighbor spin-spin correlations (or bond energies). The thickness of each bond represents the strength of the nearest-neighbor antiferromagnetic spin correlation. Circles of different colors indicate whether the spin correlation between a lattice site and the reference point (the largest light-blue circle) is positive or negative, while the size of each circle reflects the magnitude of the correlation.}
\label{fig:DMRGDelta0p0}
\end{figure*}

\begin{figure*}[htp]
  \centering
  \includegraphics[width=1.0\textwidth]{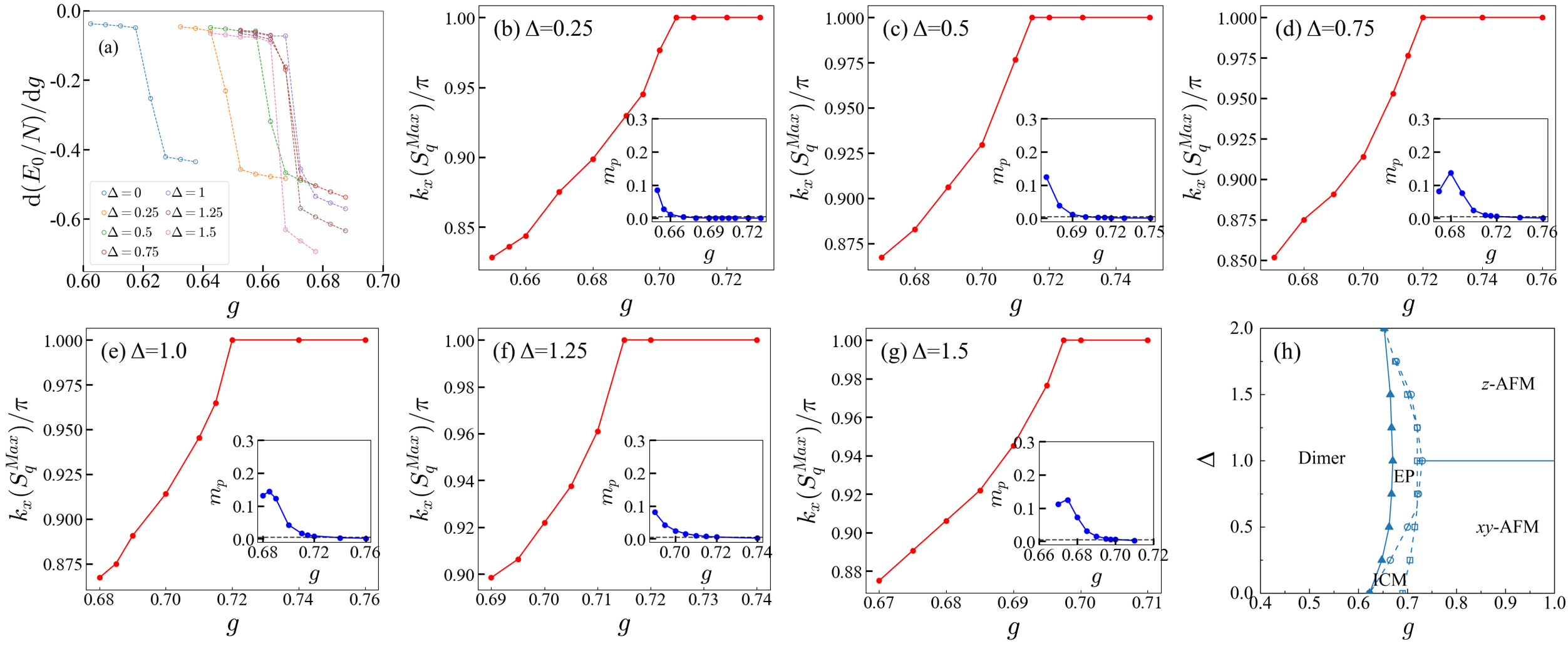}
  \caption{(a) First derivative of the ground-state energy with respect to $g$ for several values of $\Delta$. (b–g) Position of the magnetic Bragg peak in the static spin structure factor as a function of $g$; the insets show the corresponding plaquette order parameter for each $\Delta$. (h) Phase diagram obtained using DMRG on an $L_y=6$ cylinder. Solid triangles mark first-order phase transitions , identified from discontinuities in the first derivative $d(E_0/N)/dg$. Open circles are determined by a finite plaquette order threshold; the remaining open square correspond to the magnetic Bragg peak reaching the M point.}
\label{fig:DMRGOtherDelta}
\end{figure*}

\subsection{DMRG calculations with cylinders}
    
As discussed in the previous sections, the finite-cluster mean-field treatment of strong intradimer correlations in CMFT calculations is hard to get very accurate transition points between the dimer and plaquette phases using small clusters. To improve the accuracy of phase boundaries, we performed DMRG calculations on $L_x=24$, $L_y=4,6$ cylinders with open boundary along $x$ direction and periodic boundary along $y$ direction. As shown in Fig.~\ref{fig:DMRGOtherDelta}(a), the first derivative of the ground-state energy per site with respect to $g$ at different XXZ anisotropy $\Delta$ exhibits a clear jump, signalling a first-order transition. The phase boundary of Dimer-EP obatained from $L_y=4$ cylinder changes little compared with $L_y=6$, so $L_y=6$ cylinder is large enough to get the accurate phase boundary between Dimer and EP phases. For $\Delta=1$ case, the phase transition point is $g_{c1}\approx0.67$ which is very close to the point obtained by other groups with different methods.
    
For the EP-AFM phase boundary, CMFT calculations in previous subsection have yielded reliable results at $\Delta = 1$.  However, this is not the case for small $\Delta < 1$. Through careful verification using DMRG, we find that a new incommensurate magnetic (ICM) phase emerges in the small-$\Delta$ region. The presence of incommensurate magnetic order makes it difficult for small-size ED and small-cluster CMFT to accurately capture this order, leading to incorrect phase boundaries. Therefore, DMRG calculation with long cylinders is better to resolve the phase diagram in the $\Delta < 1$ regime. we adopted an open boundary condition along the $x$-direction with a sufficiently long $L_x = 24$ for our simulations.

Let us first consider the planar XX model at $\Delta=0$. The first and second derivatives of the ground-state energy, shown in Fig.~\ref{fig:DMRGOtherDelta}(a) and Fig.~\ref{fig:DMRGDelta0p0}(a), indicate a first-order phase transition from the dimer phase to another phase at a critical coupling $g_{c1}\approx0.625$. Beyond this point, no clear signature of a further phase transition is observed within the interval $g\in[0.63,0.72]$. A natural question arises: what are the phases when $g>g_{c1}$? Are there EP phase, AFM phase, or something else? To identify these possible phases, we calculate the spin structure factor $S(\mathbf{q})=\frac{1}{N}\sum_{ij}e^{i\mathbf{q}(\mathbf{r}_j-\mathbf{r}_i)}\langle S_iS_j \rangle$ at $g=0.65$, shown in Fig.~\ref{fig:DMRGDelta0p0}(b). Two bright magnetic Bragg peaks appear near the $M=(\pi,\pi)$ point. As $g$ increases, the position of these peaks shifts toward the $M$ point, as indicated by the red curve in Fig.~\ref{fig:DMRGDelta0p0}(c). In the inset, the blue curve shows that the plaquette order parameter remains nearly zero. These observations indicate that, after the first-order transition, the system enters an ICM phase. To visualize the order more directly, Fig.~\ref{fig:DMRGDelta0p0}(e) displays the bond energy and the real-space spin correlation between each site and a reference spin (marked by the largest circle) located at the center of the cylinder. The nearest-neighbor correlations (bond energies) around the center are uniform and aligned, showing no signature of a plaquette order. Furthermore, the magnitude of the spin correlations (indicated by circle size) does not decay monotonically with distance from the reference point, a clear hallmark of ICM order. For a more quantitative analysis of the real-space spin correlations, we examine their decay along the cylindrical axis ($x$-direction) which is shown in Fig.~\ref{fig:DMRGDelta0p0}
(d). To capture long-distance behavior while minimizing boundary effects, the reference point is chosen two sites away from the left edge. The decay of spin correlations along $x$ is then plotted for several values of $g$. For $g=0.63,0.65,0.67$, the correlations exhibit oscillatory, quasi-power-law decay with an incommensurate period. In contrast, for $g=0.70$ and 
$g=0.72$, the decay reverts to a typical form of N\'{e}el antiferromagnetic order in the $xy$ plane, suggesting an additional transition from the ICM phase to the commensurate $xy$-AFM phase. The point at which the magnetic Bragg peak reaches the $M=(\pi,\pi)$ point in Fig.~\ref{fig:DMRGDelta0p0}(c) can be used to define the transition $g_{c2}\approx0.69$ between the ICM phase and $xy$-AFM phase. At $g=0.70$, which is close to this transition, the real-space correlations decay exponentially (due to finite size), while at $g=0.72$ (farther into the ordered phase) they show clear quasi-long-range algebraic decay. In summary, based on the energy derivatives and the evolution of magnetic Bragg peaks, we establish three distinct phases for $\Delta=0$: a dimer phase, an ICM phase, and a $xy$-AFM phase. The absence of a peak in the second energy derivative at $g_{c2}$ [Fig.~\ref{fig:DMRGDelta0p0}(a)] is likely due to finite-size effects that may mask the incommensurate ordering.

Next we turn to the regime $0<\Delta<1$. Compared to the $\Delta=0$ case, the EP phase now competes directly with the ICM phase, as evidenced in Fig.~\ref{fig:DMRGOtherDelta}(b–d). Characterizing the transition between these phases ideally requires extrapolating their respective order parameters to the thermodynamic limit. However, such an extrapolation is challenging because DMRG calculations are hard to achieve well-converged results for cylinders with $L_y\ge8$. In lieu of a full extrapolation, we adopt a practical criterion: a plaquette order parameter $m_p\ge 5.0\times 10^{-3}$
is taken as the threshold indicating the establishment of long-range plaquette order in the thermodynamic limit. Similar techniques have also been used to determine the magnetic phase boundaries in other DMRG calculations on frustrated spin models~\cite{ZZhu2015, ZZhu2017, ZZhu2018, Maksimov2019, Gallegos2025}. For the subsequent transition from the EP phase to the $xy$-AFM phase, we continue to use the evolution of the magnetic Bragg peak, identifying the transition at the point where the peak reaches the $M$ point. For $\Delta=0.25,0.5,0.75$ shown in Fig.~\ref{fig:DMRGOtherDelta}(b-d), applying these two criteria, we map out the $0<\Delta<1$ phase boundaries obtained by DMRG, which are indicated by the light blue lines in Fig.~\ref{fig:DMRGOtherDelta}(h). 

For the regime $\Delta\ge1$, we similarly employ the plaquette order parameter threshold and the evolution of the magnetic Bragg peak to estimate the phase transition points. As shown in Fig.~\ref{fig:DMRGOtherDelta}(f) and~\ref{fig:DMRGOtherDelta}(g), These two criteria yield consistent estimates that are nearly identical. These points are included in the $\Delta\ge1$ phase diagram in Fig.~\ref{fig:DMRGOtherDelta}(h). For $\Delta=1$ shown in Fig.~\ref{fig:DMRGOtherDelta}(e), the magnetic Bragg peaks converge to the $M$ point around $g\approx0.72$, which is somewhat smaller than the phase transition point previously reported using various numerical methods (ranging from $0.77$ to $0.828$). This discrepancy suggests that finite-$L_y$ effects remain non-negligible in our DMRG calculations near $\Delta=1$, where quantum fluctuation is stronger compared to the XX limit.

\subsection{Comparison between three methods}

\begin{figure}
	\centering
	\includegraphics[width=0.5\textwidth]{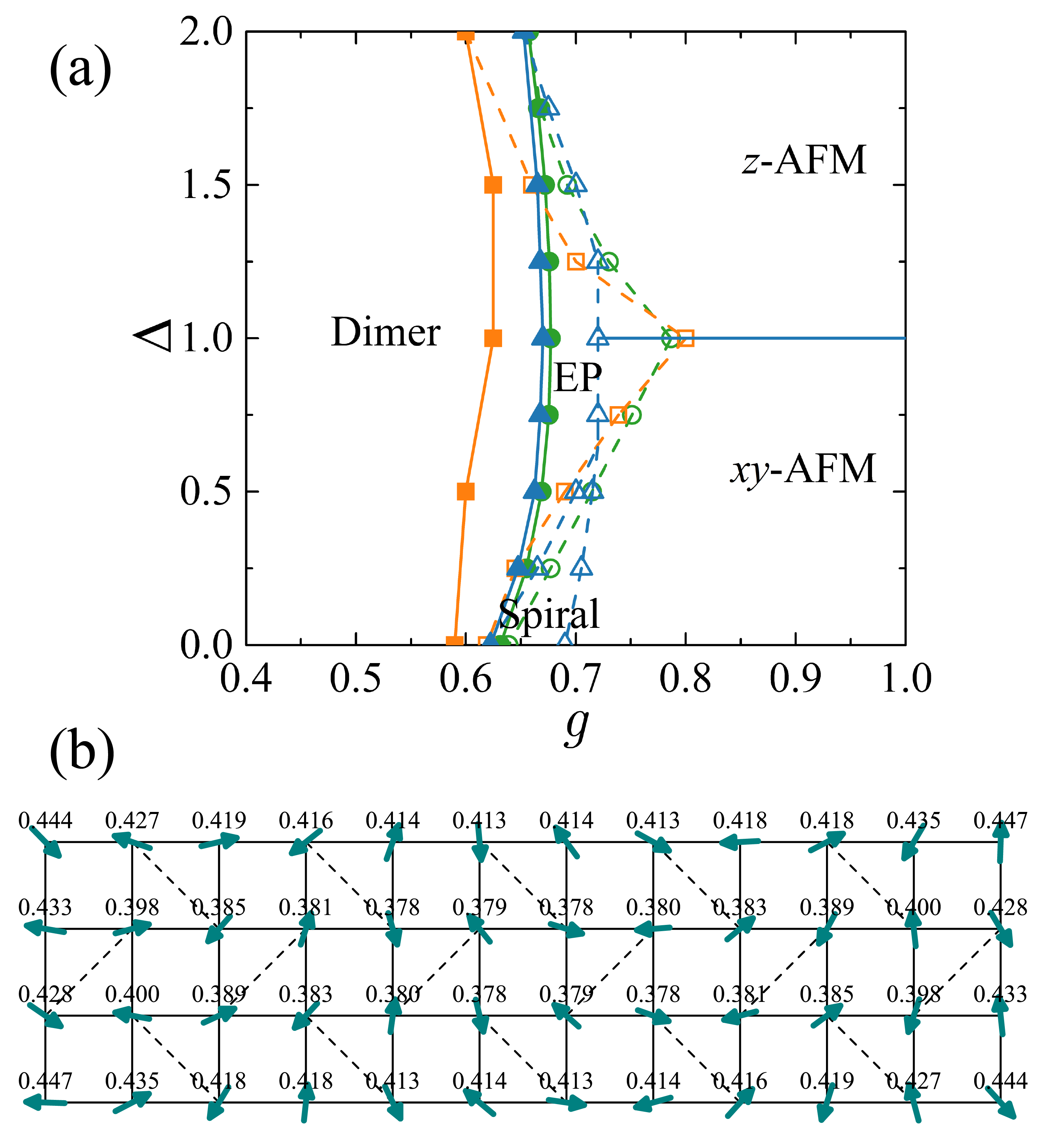}
	\caption{(a) Phase diagram obtained from ED, CMFT, and DMRG calculations, with results represented by green, orange, and blue lines, respectively. (b) The magnetic order strength on each site obtained by CMFT using 12$\times$4 cluster at $\Delta$ = 0.0, $g$ = 0.65. The magnitude of $x$ and $y$ components of the arrows represent the magnetic order strength $\langle S^x \rangle$ and $\langle S^y \rangle$ of different sites, respectively. And $\langle S^z \rangle$ are almost zero for all sites.}
	\label{fig:Phase_Full}
\end{figure}

In order to better compare the results, the phase diagrams from all three computational methods are placed together for a side-by-side comparison which is shown in Fig.~\ref{fig:Phase_Full} (a). Across all results obtained from the three methods, the intermediate EP phase shrinks as one moves away from the isotropic  $\Delta=1$ case. However, the precise point at which the EP phase disappears varies slightly between the methods. For small $\Delta$, both 32-site ED and finite $L\times L$-cluster CMFT fail to capture the ICM phase and thus cannot correctly identify the phase transitions in this regime. In contrast, DMRG on long cylinders reliably identifies the EP, ICM, and $xy$-AFM phases and their transitions. According to DMRG, at $\Delta=0$, there is a direct first-order transition from the dimer phase to the ICM phase, followed by a transition to the $xy$-AFM phase. To further confirm the coplanar spiral nature of this ICM phase, we return to CMFT calculations. Informed by the DMRG results, we employ an elongated rectangular cluster of $L_x=12$, $L_y=4$ at $g=0.65$ for our calculations. As shown in Fig.~\ref{fig:Phase_Full} (b), the spin (magnetic) order at each site is represented as a vector, displaying clear coplanar spiral characteristics with nearly equal magnitudes (indicated in the figure) across the central region, while their orientations vary continuously in a common plane. This structure arises from frustrated antiferromagnetic interactions, where neighboring spins form obtuse angles, preserving their antiferromagnetic nature. These results confirm the presence of a coplanar incommensurate spiral phase. As $\Delta$ increases, the spiral phase competes with the EP phase. The phase transition line between EP phase and spiral phase is difficult to characterize precisely with the currently accessible cylinder widths ($L_y\le6$). Similarly, determining the exact $\Delta$ at which the spiral phase vanishes faces the same limitation. Our numerical data indicate that the spiral phase vanishes at $\Delta \approx 0.72$. 

In the intermediate regime $0.75 < \Delta < 1.25$, particularly near $\Delta = 1$, the phase boundaries determined by DMRG-based on the threshold of the finite plaquette order or the magnetic Bragg peak reaching the M point-are smaller than those from other methods, which may be attributed to finite-$L_y$ effects. However, the CMFT results for the disappearance of the $xy$-AFM and $z$-AFM order parameters appear reliable, as they remain nearly consistent between $4\times4$ and $6\times6$ clusters. For $\Delta \ge 1.25$, CMFT seems to overestimate the $z$-AFM order, while ED and DMRG yield very similar results, providing a more accurate estimate for the transition from the EP phase to the $z$-AFM phase. As for the dimer-to-EP transition, DMRG gives the most reliable phase boundary. Integrating all computational results and the preceding discussion, we obtain a more robust ground-state phase diagram for the Shastry-Sutherland XXZ model, shown in Fig.~\ref{fig:lattice}(b).

In the range of $0.75<\Delta<1.25$, the region bounded on the left by the light-blue line (derived from DMRG via the shift of the magnetic Bragg peak to the $M$ point) and on the right by the orange line (marking the disappearance of AFM order from CMFT) may host either a quantum spin liquid (QSL) phase~\cite{PhysRevB.105.L060409,1655430696868-504655297,PhysRevB.105.L041115,PhysRevB.111.134411,corboz2025quantumspinliquidphase} or exhibit disorder-like behavior due to proximity to a deconfined quantum critical point (or line)~\cite{PhysRevX.9.041037,PhysRevB.105.L060409,PhysRevB.107.L220408,PhysRevLett.133.026502,li2024spinonsnewshastrysutherlandlattice,YCui2025}. We have added a red gradient‑shaded area to illustrate this possible phase region in Fig.~\ref{fig:lattice}(b). Fully confirming its presence would require substantial further efforts and faces considerable challenges, such as performing careful finite-size extrapolations of order parameters on significantly larger systems. We therefore leave the investigation of a possible QSL phase and its phase region under XXZ anisotropy for future study. In addition, the spin-liquid-like behaviors in RE$_2$Be$_2$GeO$_2$~\cite{doi:10.1021/acs.inorgchem.0c03131,PhysRevB.110.014412,li2024spinonsnewshastrysutherlandlattice,PhysRevB.110.144445} could be due to its effective interaction parameters occupying a crucial regime. This may be near the spiral phase boundary, within a possible QSL phase, or close to the deconfined quantum critical point or line.
    
\section{Summary and Discussion}
\label{sec:Summary}

In this work, we have determined the ground-state phase diagram of the $S$=1/2 XXZ model on Shastry-Sutherland lattice using a combination of exact diagonalization (ED), cluster mean-field theory (CMFT) and density matrix renormalization group (DMRG). Our numerical results reveal five distinct phases across the parameter space of anisotropy ($\Delta$) and coupling ratio ($g = J/J^\prime$): a dimer phase, a plaquette VBS phase, a $z$-AFM phase, a $xy$-AFM phase and especially a spiral phase. CMFT calculations on both E- and F-type clusters reconfirm that the intermediate plaquette phase is characterized by tetramerization within empty plaquettes, it is an EP phase. However, this method fails to accurately locate the dimer-to-EP phase transition due to its inadequate mean-field treatment of strong dimer bonds. We thus employed DMRG to precisely determine this first-order phase boundary. In contrast, the other transitions appear to be continuous. In the small $\Delta$ regime, we find and characterize a new spiral phase. Combined three numerical methods, we have established a comprehensive and reliable phase diagram. Our phase diagram supports the conjecture that the spin-liquid-like behaviors observed in RE$_2$Be$_2$GeO$_2$ originate from the competition between the EP, spiral, and $xy$-AFM phases near their boundaries. This understanding offers crucial insights into XXZ anisotropy in the Shastry-Sutherland model and provides a valuable framework for interpreting experiments on related materials.

\begin{acknowledgments}
This project is supported by NSFC-12474248, NKRDPC-2022YFA1402802, NSFC-92165204, NSFC-12494591, Guangdong Basic and Applied Basic Research Foundation (Grant No. 2023B1515120013), Guangdong Provincial Key Laboratory of Magnetoelectric Physics and Devices (No. 2022B1212010008), Guangdong Fundamental Research Center for Magnetoelectric Physics (Grant No.2024B0303390001) and Guangdong Provincial Quantum Science Strategic Initiative (Grant No.GDZX2401010).
\end{acknowledgments}

	\bibliography{XXZSSL}
	
\end{document}